\documentclass[12pt]{iopart}

\usepackage{iopams}

\DeclareSymbolFont{lettersA}{U}{pxmia}{m}{it}
\DeclareMathAlphabet{\mathsfsl}{OT1}{cmss}{m}{sl}

\SetSymbolFont{lettersA}{bold}{U}{pxmia}{bx}{it}
\DeclareFontSubstitution{U}{pxmia}{m}{it}
\DeclareSymbolFontAlphabet{\mathfrak}{lettersA}
\DeclareMathSymbol{\piup}{\mathord}{lettersA}{"19}
\DeclareMathSymbol{\iTheta}{\mathalpha}{letters}{2}

\usepackage{amssymb}
\usepackage{epstopdf}
\usepackage{graphicx}
\usepackage[hang,scriptsize]{subfigure}
\usepackage{bbm}
\usepackage{mathrsfs}
\usepackage{xcolor}
\usepackage[dvipdfm,colorlinks=true,pdfstartview=FitV,linkcolor=blue,citecolor=blue,urlcolor=blue]{hyperref}

\makeatletter

\newcommand{\Rmnum}[1]{\expandafter\@slowromancap\romannumeral #1@}
\makeatother

\newcommand{\ii}{\mathrm{i}}

\newcommand{\ntensor}[1]{\mathord{\buildrel{\lower3pt\hbox{$\scriptscriptstyle\leftrightarrow$}}\over{#1}}}

\newcommand{\blue}[1]{\textcolor{blue}{#1}}

\begin{document}

%%%%%%%%%%%%%
%%% TITLE %%%
%%%%%%%%%%%%%

\title{Deep subwavelength beam propagation in extremely loss-anisotropic metamaterials}

\author{Yingran He$^{1,2}$, Lei Sun$^{1}$,
    Sailing He$^{2}$, and Xiaodong Yang$^{1}$}

\address{
    $^{1}$Department of Mechanical and Aerospace Engineering, \\
    Missouri University of Science and Technology,
    Rolla, MO 65409, USA}

\address{
    $^{2}$Centre for Optical and Electromagnetic Research, \\
    Zhejiang Provincial Key Laboratory for Sensing Technologies, \\
    Zhejiang University, Hangzhou 310058, China}

\ead{yangxia@mst.edu}

\begin{abstract}
Metal-dielectric multilayer metamaterials with extreme loss-anisotropy,
in which the longitudinal component of the permittivity tensor has ultra-large
imaginary part, are proposed and designed.
Diffraction-free deep subwavelength beam propagation and manipulation,
due to the nearly flat iso-frequency contour (IFC), is demonstrated in such
loss-anisotropic metamaterials.
It is also shown that deep subwavelength beam propagation can be realized
in practical multilayer structures with large multilayer period, when the
nonlocal effect is considered.
\end{abstract}


\maketitle

%%%%%%%%%%%%%%%%%
%%% MAIN TEXT %%%
%%%%%%%%%%%%%%%%%

%%%%%% INTRODUCTION %%%%%%%%%%%%%%%%%%%%%%%%%%%%%%%%%%%%%%%%%%%%%%%%%%%%%%%%%
\section{Introduction}
Light beam with subwavelength confinement and propagation beyond diffraction limit
is highly desirable for many optical integration applications.
To enable the subwavelength optical confinement, optical materials supporting ultra-high
wave vectors are necessary.
Since the refractive indices of natural materials are quite limited at optical frequency,
metamaterials with artificially engineered subwavelength meta-atoms are designed to exhibit
ultra-high refractive indices so as to achieve large wave vectors \cite{Shin2009PRL,Choi2011Nat}.
To obtain diffraction-free deep subwavelength beam propagation, a nearly flat IFC curve over
a broad range in $k$-space is required, so that all spatial components will propagate with
the same phase velocity along the longitudinal direction \cite{Han2008NanoLett}.
Extremely anisotropic metamaterials with infinite real part of permittivity have been theoretically
proposed to achieve flat IFC and consequently the subwavelength beam propagation without
diffraction \cite{Catrysse2012CLEO,Catrysse2011PRL}.

%%---%%
Recently, metal-dielectric multilayer metamaterials with indefinite permittivity tensor
have been utilized to demonstrate intriguing applications of
negative refraction \cite{Yao2008Sci},
subwavelength imaging \cite{Liu2007Sci},
enhanced photonic density of states \cite{Krishnamoorthy2012Sci},
and broadband light absorbers \cite{Cui2012NanoLett},
together with ultra-high refractive indices for subwavelength optical waveguides \cite{He2012JOSAB}
and indefinite cavities \cite{Yang2012NatPhoto}.
In the present work, we propose the concept of extreme loss-anisotropy
in metal-dielectric multilayer metamaterials, where the longitudinal
component of the permittivity tensor has ultra-large imaginary part.
Diffraction-free deep subwavelength beam propagation and manipulation
is demonstrated in such loss-anisotropic metamaterial, due to the nearly flat IFC.

%%%%%% THEORY AND DISCUSSION %%%%%%%%%%%%%%%%%%%%%%%%%%%%%%%%%%%%%%%%%%%%%%%%%%%%%%%%%
\section{Theory and Discussion}
The inset of Fig.~\blue{1} shows the metal-dielectric multilayer structure,
where titanium oxide ($\mathrm{Ti}_{3}\mathrm{O}_{5}$) with dielectric constant
of $5.83$ and silver (Ag) are chosen \cite{Rho2010NatCom}.
When the thickness of each layer is infinitely small, the multilayer structure
can be regarded as an anisotropic effective medium with permittivity tensor of
\begin{equation}
\label{eq:emt}
    \varepsilon_{x} = f_{d}\varepsilon_{d}+f_{m}\varepsilon_{m},\quad
    \varepsilon_{y} = \left(f_{d}/\varepsilon_{d}
        + f_{m}/\varepsilon_{m}\right)^{-1},
\end{equation}
where $\varepsilon_{d}$ and $\varepsilon_{m}$ are the permittivity of titanium oxide
and silver, $f_{d}$ and $f_{m}$ ($f_{d}+f_{m}=1$) are the filling ratios of titanium
oxide and silver, respectively.
The silver filling ratio $f_{m}$ is 0.45, and its permittivity is from the experimental
results \cite{Johnson1972PRB}.
The dependence of permittivity tensor on wavelength $\lambda_{0}$ is shown in Fig.~\blue{1(a)},
which is calculated using Eq.~(\ref{eq:emt}) according to the effective medium theory.
It is found that the longitudinal permittivity component $\varepsilon_{y}$ shows
a strong resonance at $\lambda_{0}=406.1\,\mathrm{nm}$ (at position \Rmnum{2}), where
$\mathrm{Im}(\varepsilon_{y})$ has a peak more than 230, while $\mathrm{Re}(\varepsilon_{y})$
flips its sign quickly across the resonance from negative maximum (at position \Rmnum{1})
to positive maximum (at position \Rmnum{3}).
Previously, ultra-large $\mathrm{Re}(\varepsilon_{y})$ (at position \Rmnum{3}) has been
utilized to realize subwavelength beam propagation without
diffraction \cite{Catrysse2012CLEO,Catrysse2011PRL}.
However, the behavior of ultra-high $\mathrm{Im}(\varepsilon_{y})$ has not been considered before.
It is intuitively thought that a large $\mathrm{Im}(\varepsilon_{y})$ will increase the beam propagation loss.
On the contrary, it will be demonstrated that the extreme loss-anisotropy with ultra-large
imaginary part can enable low-loss diffraction-free subwavelength beam propagation.
For TM-polarized light with non-vanishing $E_{x}$, $E_{y}$ and $H_{z}$ field components,
the corresponding IFC is determined by
\begin{equation}
\label{eq:ifc}
    k_{x}^{2}/\varepsilon_{y} + k_{y}^{2}/\varepsilon_{x} = k_{0}^{2},
\end{equation}
where $k_{x}$ is the transverse $k$-vector and $k_{y}$ is the propagation $k$-vector.
Figure~\blue{1(b)} shows that an ultra-flat IFC over a large $k_{x}$ range is supported
at the resonance wavelength $\lambda_{0}=406.1\,\mathrm{nm}$ (position \Rmnum{2}),
with $\varepsilon_{x}=1.06+0.098\ii$ and $\varepsilon_{y}=10.48+231.70\ii$.
The propagation $k$-vector $\mathrm{Re}(k_{y})$ remains a constant for all different $k_{x}$.
While the imaginary part of the propagation $k$-vector $\mathrm{Im}(k_{y})$ is quite small.
To understand the behavior of the flat IFC, Eq.~(\ref{eq:ifc}) can be rewritten as
\begin{equation}
\label{eq:approx-ifc}
    k_{y} = \sqrt{\varepsilon_{x}\left( k_{0}^{2} - \frac{k_{x}^{2}}{\varepsilon_{y}} \right)}
        \approx \sqrt{\varepsilon_{x}}k_{0} - \sqrt{\varepsilon_{x}}\frac{k_{x}^{2}}{2k_{0}\varepsilon_{y}},
\end{equation}
where the approximation $\left|\varepsilon_{y}\right|\gg1$ has been used to derive the above formula.
Since $\varepsilon_{x}$ is dominated by its real part, and $\varepsilon_{y}$ is dominated by its
imaginary part at the resonance wavelength, $\mathrm{Re}(k_{y})\approx\sqrt{\varepsilon_{x}k_{0}}$
and $\mathrm{Im}(k_{y})\approx\sqrt{\varepsilon_{x}}k_{x}^{2}/\left( 2k_{0}\mathrm{Im}(\varepsilon_{y})\right)$,
which shows that $\mathrm{Re}(k_{y})$ is independent of $k_{x}$ and $\mathrm{Im}(k_{y})$ is weakly
proportional to $k_{x}^{2}$.
Moreover, the propagation loss $\mathrm{Im}(k_{y})$ is inversely proportional to $\mathrm{Im}(\varepsilon_{y})$
and the large material loss $\mathrm{Im}(\varepsilon_{y})$ actually enables the low-loss beam propagation.
The ultra-large $\mathrm{Im}(\varepsilon_{y})$ in extremely loss-anisotropic metamaterial not only gives
rise to an ultra-flat IFC over a broad $k_{x}$ range, but also results in ultra-small beam propagation loss.

%%---%%
To further illustrate the importance of large $\mathrm{Im}(\varepsilon_{y})$, the comparisons of IFC curves
at three different wavelengths of \Rmnum{1}, \Rmnum{2} and \Rmnum{3} are shown in Fig.~\blue{1(c)},
which corresponds to negatively maximized $\mathrm{Re}(\varepsilon_{y})$, maximized $\mathrm{Im}(\varepsilon_{y})$
and maximized $\mathrm{Re}(\varepsilon_{y})$, respectively.
It indicates that the maximized $\mathrm{Im}(\varepsilon_{y})$ case results in the flattest IFC
with almost zero curvature, while the other two IFCs show positive and negative curvatures, respectively.
These behaviors can be clearly understood using Eq.~(\ref{eq:approx-ifc}), in which a purely real
$\varepsilon_{y}$ will contribute to $\mathrm{Re}(k_{y})$ and influence the curvature of the IFC
(a positive $\varepsilon_y$ leads to a negative curvature and vice versa).
It is clear that the maximized $\mathrm{Im}(\varepsilon_{y})$ case with zero curvature will achieve
diffraction-free subwavelength beam propagation.

%%---%%
Next, diffraction-free deep-subwavelength beam propagation will be demonstrated
from the numerical simulation based on finite element method (FEM).
Here it is worthwhile to define the minimal waist size of light beam which can
propagate inside the multilayer structure without diffraction.
It is known from Eq.~(\ref{eq:ifc}) that $k_{x}^{2}/\varepsilon_{y}\ll k_{0}^{2}$
is required to obtain a nearly flat IFC, then the minimal beam waist size $w_{\mathrm{min}}$
turns out to be $2\pi/\mathrm{max}(k_{x})=\lambda_{0}\sqrt{\left|\varepsilon_{y}\right|}$.
Figure~\blue{2} shows the propagation of ultra-narrow Gaussian beams
(with a waist size of $40\,\mathrm{nm}\sim0.1\lambda_{0}>w_{\mathrm{min}}$) inside the
loss-anisotropic metamaterials with different geometries at the resonance
wavelength $\lambda_{0}=406.1\,\mathrm{nm}$.
The effective medium results and the realistic multilayer results
are shown in Fig.~\blue{2(a--c)} and Fig.~\blue{2(d--f)}, respectively.
Figures~\blue{2(a)} and \blue{2(d)} show that ultra-narrow Gaussian beams
can propagate over a long distance without any wave-front distortion.
Two subwavelength beams with $150\,\mathrm{nm}$ center-to-center distance
remain well-defined as the beams propagating across the multilayer from the
bottom to the top.
It is emphasized that the subwavelength beam confinement is entirely due to
the unique loss-anisotropic property, and the beam path is solely determined
by the launching location.
This is distinguished from the situation in a subwavelength waveguide,
where the mode is confined by the waveguide boundary.

%%---%%
Besides the straight beam propagation, the flow of light can be flexibly modeled
through controlling the local metamaterial properties.
For instance, the beam path can be manipulated by gradually varying the direction
of multilayers, since the direction of beam propagation is always vertical to the
multilayer interface.
The designed geometries for achieving $90^{\circ}$ and $180^{\circ}$ bending of
subwavelength beams are shown in Figs.~\blue{2(b--c)} and \blue{2(e--f)}, for both
the effective media and the multilayer structure.
In the effective medium calculation, the anisotropic permittivity tensor
depends on the tilted angle of the multilayer interface.
In the local coordinate $(u, v)$, the components of the permittivity tensor can still
be determined by the mixing formula in Eq.~(\ref{eq:emt}).
The permittivity tensor expression in the global coordinate $(x, y)$ is related to that
in the local coordinate as
\begin{equation}
\label{eq:to}
    \ntensor{\varepsilon}(x,y) = \left(
         \begin{array}{cc}
            \varepsilon_{u}\cos^{2}\theta + \varepsilon_{v}\sin^{2}\theta
                & (\varepsilon_{u} - \varepsilon_{v})\sin\theta\cos\theta \\
            (\varepsilon_{u} - \varepsilon_{v})\sin\theta\cos\theta
                & \varepsilon_{u}\sin^{2}\theta + \varepsilon_{v}\cos^{2}\theta
         \end{array}\right),
\end{equation}
where $\theta$ is the local tilted angle of multilayer with respect to the $+x$ axis,
$\varepsilon_{u}$ and $\varepsilon_{v}$ are the local permittivity tensor components
along and normal to the multilayer, respectively.
The simulation results indicate that the flow of light can indeed be manipulated while
maintaining the deep-subwavelength beam confinement and diffraction-free propagation.
For the results of $180^{\circ}$ bending shown in Figs.~\blue{2(c)} and \blue{2(f)},
it is noted that the wave-front becomes tilted at the output section (but the energy
flow is still vertical to the interface as a result of the flat IFC).
This is due to the fact that the light traveling at the inner side undergoes less
optical path than the light traveling at the outer side.
The phase difference arising from the light path difference leads to the beam wave-front tilting.

%%---%%
Figure~\blue{2} shows that the multilayer structure simulation results agree very well
with the EMT results, indicating that the multilayer structure with period $a=20\,\mathrm{nm}$
($f_{m}=0.45$) can represent the loss-anisotropic effective medium well.
However, the fabrication of such thin layers is very challenging in reality
(but possible \cite{Chen2010OE}).
It will be interesting to study the property of metal-dielectric multilayer with a large period $a$,
where the nonlocal effect has to be taken into account \cite{Elser2007APL}.
The dispersion relation describing the realistic multilayer structure for TM-polarized light is
\begin{equation}
\label{eq:pc}
    \cos\left[k_{y}(a_{m}+a_{d})\right] = \cos(k_{m}a_{m})\cos(k_{d}a_{d})
        -\gamma_{\scriptscriptstyle\mathrm{TM}}\sin(k_{m}a_{m})\sin(k_{d}a_{d}),
\end{equation}
which is derived by treating the layered structure as a one-dimensional photonic crystal.
Here $\gamma_{\scriptscriptstyle\mathrm{TM}}=\left(\varepsilon_{d}k_{m}/\varepsilon_{m}k_{d}
+ \varepsilon_{m}k_{d}/\varepsilon_{d}k_{m}\right)/2$,
$k_{m}=\sqrt{\varepsilon_{m}k_{0}^{2}-k_{x}^{2}}$,
and $k_{d}=\sqrt{\varepsilon_{d}k_{0}^{2}-k_{x}^{2}}$.
The IFCs corresponding to the realistic multilayer structure with $a=40\,\mathrm{nm}$
are shown in Fig.~\blue{3(a)}.
It is found that the IFC curve at $\lambda_{0}=406.1\,\mathrm{nm}$ is no longer flat
due to the nonlocal effect.
That is to say, the effective permittivity tensor becomes strongly wave vector dependent,
so that the permittivity components $\varepsilon_{x}$ and $\varepsilon_{y}$ will be
functions of not only the frequency but also the $k$-vector.
The frequency corresponding to the flattest IFC will then be shifted.
For the multilayer structure with $a=40\,\mathrm{nm}$, it turns out that the flattest
IFC occurs at $\lambda_{0}=418.3\,\mathrm{nm}$ (which can be mathematically determined
by finding the working wavelength with zero IFC curvature), as shown in Fig.~\blue{3(a)}.
The propagation of subwavelength Gaussian beams at the two wavelengths in the multilayer
structure with $a=40\,\mathrm{nm}$ are shown in Fig.~\blue{3(b)}.
As can be expected from the IFC curves in Fig.~\blue{3(a)}, the Gaussian beams at
$\lambda_{0}=406.1\,\mathrm{nm}$ suffer strong diffraction, resulting in distorted beam profiles.
In comparison, the beam profiles remain well-defined for the Gaussian beams at $\lambda_{0}=418.3\,\mathrm{nm}$.

%%%%%% CONCLUSION %%%%%%%%%%%%%%%%%%%%%%%%%%%%%%%%%%%%%%%%%%%%%%%%%%%%%%%%%
\section{Conclusion}
In conclusion, extremely loss-anisotropic metamaterial is designed using metal-dielectric
multilayer structures.
The IFC corresponding to such metamaterial turns out to be ultra-flat over a broad $k$-vector range.
This unique property is then utilized to obtain diffraction-free deep subwavelength beam propagation.
Furthermore, it is shown that the propagation of light beams can be manipulated flexibly by tuning
the direction of multilayer structure.
Moreover, the nonlocal effect occurring in multilayer structures with large multilayer period is
investigated.
It is found that diffraction-free beam propagation is still possible after taking into account
the nonlocal effect.
The current study is very attractive for many applications such as optical imaging, optical integration,
on-chip optical communication.

%%%%%% ACKNOWLEDGEMENT %%%%%%%%%%%%%%%%%%%%%%%%%%%%%%%%%%%%%%%%%%%%%%%%%%%%%%
\section*{Acknowledgment}
This work was partially supported by the Department of Mechanical and Aerospace
Engineering and the Intelligent Systems Center at Missouri S\&T, the University
of Missouri Research Board, the Ralph E. Powe Junior Faculty Enhancement Award,
and the National Natural Science Foundation of China (61178062 and 60990322).

Y. He and L. Sun contributed equally to this work.

%%%%%% REFERENCES %%%%%%%%%%%%%%%%%%%%%%%%%%%%%%%%%%%%%%%%%%%%%%%%%%%%%%%%%%%

%%%%%%%%%%%%%%%%%%%%%%%%%%%%%%%%%%%%%%%%%%%%%%%%%%%%%%%%%%%%%%%%%%%%%%%%%%%%%

%%%%%%%%%%%%%%%%%%%%%%%%%
%%% Figure Captions %%%
\newpage
\section*{Figure Captions}
%%%%%%%%%%%%%%%%%%%%%%%%%
\noindent
\textbf{Figure~1}:
    (Color online)
    (a) The dependence of the permittivity tensor on the wavelength $\lambda_{0}$ for the metal-dielectric
    multilayer structure with metal filling ratio $f_{m}=0.45$.
    $\mathrm{Re}(\varepsilon_{x})$ has been scaled up by $10$ times.
    The three Roman numbers (\Rmnum{1}, \Rmnum{2}, \Rmnum{3}) indicate the wavelengths corresponding to
    negatively maximized $\mathrm{Re}(\varepsilon_{y})$,
    maximized $\mathrm{Im}(\varepsilon_{y})$ and maximized $\mathrm{Re}(\varepsilon_{y})$, respectively.
    Inset shows the multilayer structure with $\mathrm{Ti}_{3}\mathrm{O}_{5}$ (green color)
    and silver (yellow color).
    (b) The IFC at the resonance wavelength $\lambda_{0}=406.1\,\mathrm{nm}$ (at position \Rmnum{2}).
    $\mathrm{Re}(k_{y})$ and $\mathrm{Im}(k_{y})$ are represented by the blue line and green line, respectively.
    Although $\mathrm{Im}(k_{y})$ grows with transverse wave vector $k_{x}$ (ranges from $0.0476k_{0}$
    to $0.2674k_{0}$ as $k_{x}$ increases from $0$ to $10k_{0}$), it remains small due to the large
    magnitude of $\varepsilon_{y}$, which gives a low propagation loss.
    The dashed black circle is the IFC of air.
    (c) The IFCs for the three wavelengths indicated in (a).
    The flattest IFC is obtained at the resonance wavelength with maximized $\mathrm{Im}(\varepsilon_{y})$.
\vspace{3.0mm}

\noindent
\textbf{Figure~2}:
    (Color online)
    Diffraction-free deep subwavelength beam propagation in extremely loss-anisotropic metamaterials.
    The distributions of magnetic field $H_{y}$ are shown in (a--c) for ideal effective medium and (d--f)
    for multilayer structure with period $a=20\,\mathrm{nm}$.
    (a, d) Straight beam propagation.
    (b, e) $90^{\circ}$ beam bending.
    (c, f) $180^{\circ}$ beam bending.
    The center-to-center distance of the two beams is $150\,\mathrm{nm}$ in all the simulations.
\vspace{3.0mm}

\noindent
\textbf{Figure~3}:
    (Color online)
    (a) The IFCs for realistic multilayer structure with $a=40\,\mathrm{nm}$ at two
    different wavelengths.
    (b) The Gaussian beams propagation at the two wavelengths.
    Diffraction-free beam propagation is achieved after taking into account the nonlocal
    effect induced wavelength shift.

%%%%%%%%%%%%%%%%%%%%%%%%%%%%%%%%%%%%%
%%%% Figures %%%
%%%%%%%%%%%%%%%%%%%%%%%%%%%%%%%%%%%%%
\newpage
\begin{figure}[htb]
\label{fig:fig-1}
\centerline{\includegraphics[height=5.5cm]{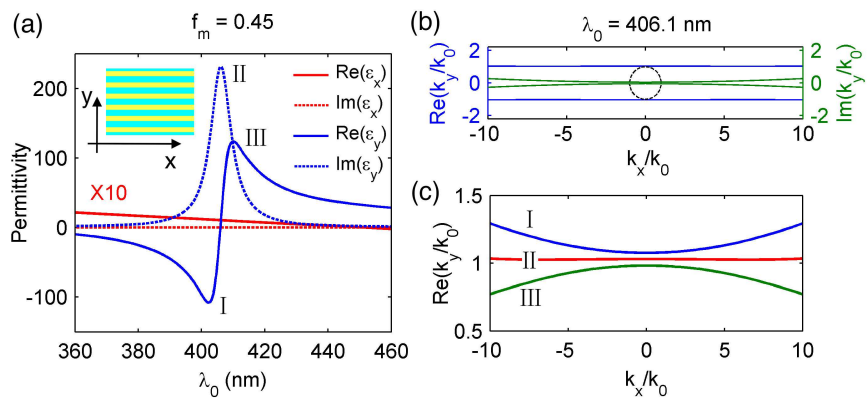}}
\caption{Yingran He, Lei Sun, Sailing He, and Xiaodong Yang}
\end{figure}

\newpage
\begin{figure}[htb]
\label{fig:fig-2}
\centerline{\includegraphics[height=8.5cm]{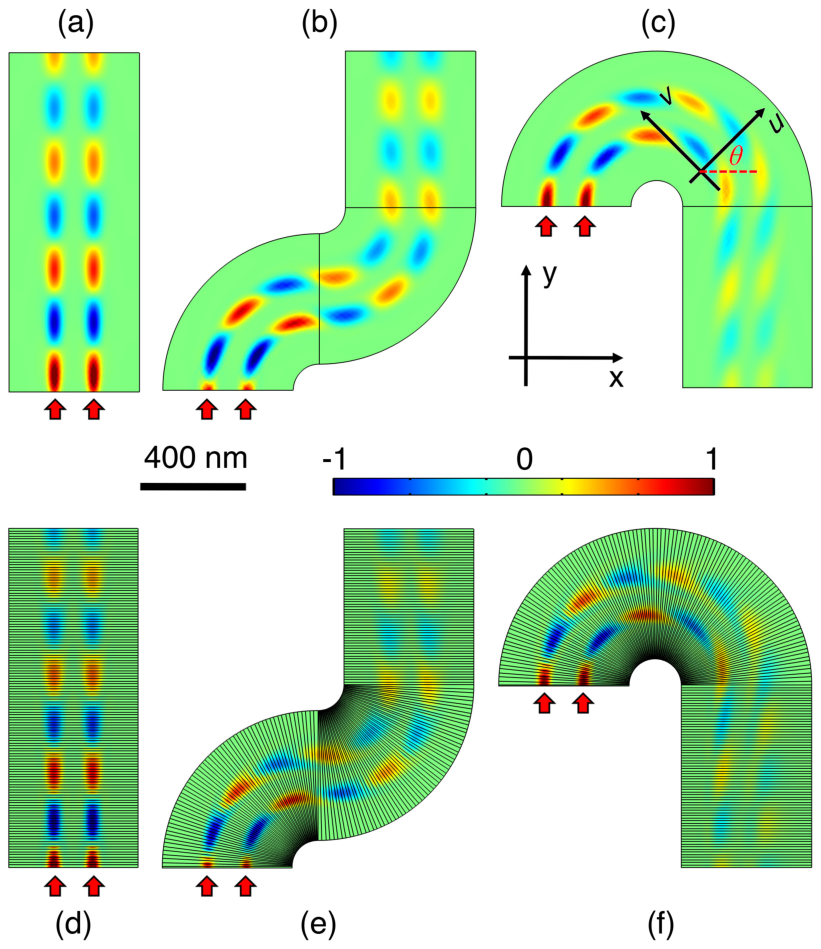}}
\caption{Yingran He, Lei Sun, Sailing He, and Xiaodong Yang}
\end{figure}

\newpage
\begin{figure}[htb]
\label{fig:fig-3}
\centerline{\includegraphics[height=5.5cm]{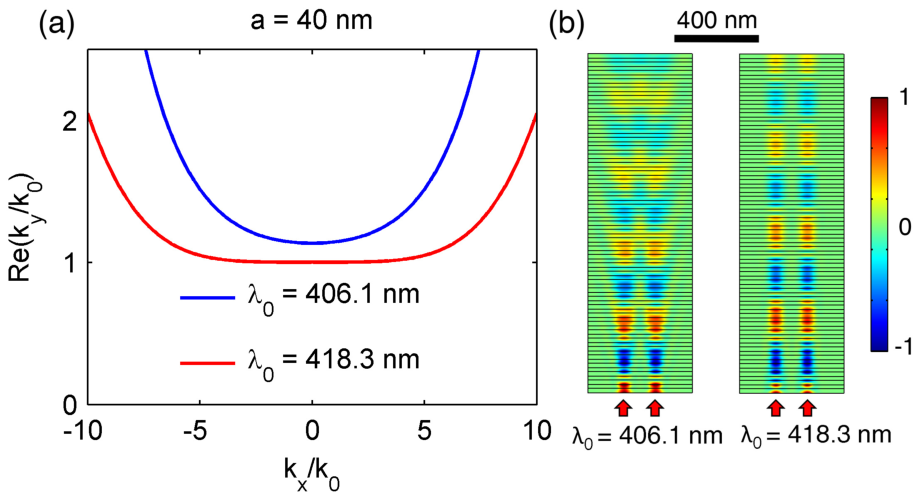}}
\caption{Yingran He, Lei Sun, Sailing He, and Xiaodong Yang}
\end{figure}

\end{document}